\documentclass{JHEP3}

\usepackage{epsfig}
\usepackage{cite}

\newcommand{\yijk}{y_{ij,k}}
\newcommand{\xija}{x_{ij,a}}
\newcommand{\xm}{x_-}
\newcommand{\xp}{x_+}
\def\li#1{\mbox{Li}_#1}

\title{\bf Assault on the NLO Wishlist: $pp \rightarrow t \bar t b \bar b$}

\author{
  G. Bevilacqua$^{a}$, M. Czakon$^{b}$,
  C. G. Papadopoulos$^{a}$, R. Pittau$^{c}$ and M. Worek$^{d,e}$
  \\ \\
  $^a$~{Institute of Nuclear Physics, NCSR Demokritos,
    GR-15310 Athens, Greece}
  \\
  $^b$~{Institut f\"ur Theoretische Physik E, RWTH Aachen University
    D-52056 Aachen, Germany}
  \\
  $^c$~{Departamento de F\'{i}sica Te\'orica y del Cosmos\\
    Centro Andaluz de F\'{i}sica de Part\'{i}culas
    Elementales (CAFPE)\\
    Universidad de Granada, E-18071 Granada, Spain}
  \\
  $^d$~{Fachbereich C, Bergische Universit\"at Wuppertal,  D-42097 Wuppertal,
    Germany}
  $^e$~{Department of Field Theory and Particle Physics,
    Institute of Physics, \\
    University of Silesia, Uniwersytecka 4, PL-40007 Katowice,
    Poland}}

\abstract{
  We present the results of a next-to-leading order calculation of 
  QCD corrections to the production of an on-shell top-anti-top quark 
  pair in association with two flavored b-jets. Besides studying the 
  total cross section and its scale dependence, we give several differential
  distributions. Where comparable, our results agree with a previous
  analysis. While the process under scrutiny is of major relevance for 
  Higgs boson searches at the LHC, we use it to demonstrate the ability 
  of our system built around \textsc{Helac-Phegas} to tackle complete 
  calculations at the frontier of current studies for the LHC. On the 
  technical side, we show how the virtual corrections are efficiently 
  computed with \textsc{Helac-1Loop}, based on the OPP method and the
  reduction code \textsc{CutTools}, using reweighting and Monte Carlo over 
  color configurations and polarizations.  As far as the real corrections 
  are concerned, we use the recently published \textsc{Helac-Dipoles} 
  package. In connection with improvements of the latter, we give the 
  last missing integrated dipole formulae necessary for a complete 
  implementation of phase space restriction dependence in the massive 
  dipole subtraction formalism.}

\preprint{PITHA 09/18 \\ WUB/09-09}

\begin{document}


\section{Introduction}
\label{sec:introduction}

Rare are processes, which received more attention
than top quark pair production in various configurations at hadron
colliders. Indeed, there is astonishing recent progress in
next-to-leading order (NLO) \cite{Czakon:2008ii,Melnikov:2009dn} and
next-to-next-to leading order 
(NNLO) \cite{Czakon:2007ej,Czakon:2007wk,Korner:2008bn,Czakon:2008zk,Bonciani:2008az,Anastasiou:2008vd,Kniehl:2008fd,Bonciani:2009nb}
calculations, as well as next-to-next-to-leading-log resummations
(NNLL) \cite{Czakon:2008cx,Beneke:2009rj,Czakon:2009zw} 
for inclusive production. For this
publication, however, we will mostly be interested in more exclusive
channels. The list for the latter is just as impressive: NLO QCD
corrections for the $t\bar{t}H$ signal
\cite{Beenakker:2001rj,Reina:2001bc,Reina:2001sf,Beenakker:2002nc,
Dawson:2002tg,Dawson:2003zu} and the
backgrounds from $t\bar{t}j$ \cite{Dittmaier:2007wz,Dittmaier:2008uj},
$t\bar{t}Z$ \cite{Lazopoulos:2008de}, $t\bar{t}b\bar{b}$
\cite{Bredenstein:2008zb,Bredenstein:2009aj}, and most recently
$t\bar{t}\gamma$ \cite{PengFei:2009ph}.

The present study pertains to the $t\bar{t}b\bar{b}$ final state. As
phenomenological motivation, we stress its relevance as irreducible
background for light Higgs boson searches in the $t\bar{t}H$ channel, when
the Higgs boson decays into a pair of b quarks. Indeed, realistic
experimental analyses of this channel \cite{atlas} including showering effects,
b-tagging efficiencies and suitable cuts to reduce multijet
backgrounds (among others $t\bar{t}jj$ with light jets misidentified as b-jets),
show a substantial smearing of what would be a sharp Higgs resonance
peak in the distribution of the invariant mass  of the $b\bar{b}$ system. 
In consequence, the knowledge of the backgrounds becomes crucial
to claim discovery or exclusion. The dominant background is obviously
direct production of the final state without resonances, {\it i.e.}
the QCD generated process $pp \rightarrow t\bar{t}b\bar{b}+X$. As
mentioned above, the NLO computation has been completed only very
recently in \cite{Bredenstein:2009aj}. Our first goal is to confirm
the results of that publication.

The second aim is to demonstrate the power of our system based on
\textsc{Helac-Phegas} \cite{Kanaki:2000ey,Kanaki:2000ms,Cafarella:2007pc}, 
\textsc{Helac-1Loop} \cite{vanHameren:2009dr}, 
\textsc{CutTools} \cite{Ossola:2006us,Ossola:2007ax} 
and \textsc{Helac-Dipoles} \cite{Czakon:2009ss} in a
realistic computation with six external legs and massive partons. This
is the first calculation at this level of complexity using the OPP reduction 
technique \cite{Ossola:2006us,Draggiotis:2009yb} and modern
unitarity based methods \cite{Bern:1994zx,Bern:1994cg,Witten:2003nn,
Britto:2004nc,Brandhuber:2005jw,Britto:2006sj,Bern:2007dw}. 
Indeed, computations with six
external legs have only been attempted with massless 
partons \cite{Ellis:2009zw,Berger:2009zg,Berger:2009ep},
whereas massive partons have only appeared in a more modest setting in
\cite{Melnikov:2009dn}. 
It is fair to say, that traditional, reduction based methods,
have proven more efficient in producing complete results until this
point (see some recent examples \cite{Dittmaier:2007th,Campbell:2007ev,Hankele:2007sb,Campanario:2008yg,FebresCordero:2008ci,Campbell:2008hh,Cordero:2009kv,Bredenstein:2009aj,Jager:2009xx}).

In principle, the publication \cite{vanHameren:2009dr} has shown that the
difficult virtual corrections can be computed within the
\textsc{Helac-1Loop}/\textsc{CutTools} framework for a multitude of 
processes of practical interest,
and $t\bar{t}b\bar{b}$ in particular. Nevertheless, it is one thing to
compute a value for a single phase space point, and another to provide
arbitrary differential distributions. Here we wish to convince the
skeptics that the latter exercise can be mastered as well. As our
title suggests, we are confident that we can now tackle any calculation
from the ``NLO Wishlist'' \cite{Bern:2008ef}.

Besides the virtual corrections, the calculation of the real radiation
contributions is not an easy task either. Despite the announcement of
several automates for Catani-Seymour \cite{Catani:1996vz,Catani:2002hc} 
dipole subtraction
\cite{Gleisberg:2007md,Seymour:2008mu,Hasegawa:2008ae,Frederix:2008hu,
Czakon:2009ss}, 
the only complete\footnote{By complete we understand phase space
  integration of subtracted real radiation and integrated dipoles in both
  massless and massive cases.} {\it and} publicly available tool is 
\textsc{Helac-Dipoles} \cite{Czakon:2009ss} (besides possibly \textsc{Sherpa}
\cite{Gleisberg:2007md}, which should become publicly available 
soon\footnote{Private communication with T. Gleisberg.}). 
In this work, we slightly increase its flexibility by
allowing for phase space restriction \cite{Nagy:1998bb,Nagy:2003tz} in the
integrated dipoles for the general massive case (most formulae were
known from \cite{Campbell:2004ch,Campbell:2005bb}).

This publication is structured as follows. In the next section, we
discuss the details of the techniques used for the evaluation of the
virtual and real corrections. Subsequently, we give the results for
the total cross sections and differential distributions, and conclude
the main part of the text. The Appendix contains the formulae of the
massive dipole formalism with phase space restriction.


\section{Technicalities}
\label{sec:technicalities}


\subsection{Virtual corrections}

For the calculation of virtual corrections we use \textsc{
Helac-1Loop}~\cite{vanHameren:2009dr}, namely the merging 
of \textsc{Helac}~\cite{Kanaki:2000ey,Kanaki:2000ms,Cafarella:2007pc} and 
the OPP \cite{Ossola:2006us,Draggiotis:2009yb} reduction  
code \textsc{CutTools} \cite{Ossola:2007ax}.  In order to further improve the
performance and speed of the system, we make use of color,
helicity and event sampling methods.

The treatment of the color degrees of freedom in \textsc{Helac} is 
based on the color-connection representation of the amplitude. A
generic QCD amplitude composed by $n_g$ gluons and $n_q$ quarks (and
of course $n_q$ antiquarks) plus possible colorless particles can be
written as
\begin{equation}
{\cal M}^{i_1\ldots,i_{n_q},a_1,\ldots,a_{n_g}}_{j_1\ldots,j_{n_q}} \; ,
\end{equation}
where $i$ and $j$ are color indices belonging to the fundamental
representation of the gauge group, whereas $a$ belongs to the adjoint.
Multiplying with $t^a_{i,j}$ and summing over $a$, for each gluon,
we end up with a uniform representation, namely
\begin{equation}\label{colorrep}
{\cal M}^{i_1\ldots,i_{n_q+n_g}}_{j_1\ldots,j_{n_q+n_g}} \; .
\end{equation}
As it is well known, the amplitude can now be (color-)decomposed as
follows
\begin{equation}\label{colordec}
{\cal M}_{j_1,j_2,\ldots,j_k}^{i_1,i_2,\ldots,i_k}=\sum_\sigma
\delta_{i_{\sigma_{1}},j_1} \delta_{i_{\sigma_{2}},j_2} \ldots
\delta_{i_{\sigma_{k}},j_k} A_\sigma \; ,
\end{equation}
with $k=n_q+n_g$, and $\sigma$ denoting a permutation of the set
$\{1,\ldots,k\}$. The Feynman rules that allow the calculation of all
$A_\sigma$ color-stripped amplitudes in the color-connection
representation have been described in 
Refs.~\cite{Kanaki:2000ey,Kanaki:2000ms,Cafarella:2007pc}. The objects we
are interested in, are the squared matrix element
\begin{equation}
\sum_{\{i\},\{j\}}|{\cal
M}_{j_1,j_2,\ldots,j_k}^{i_1,i_2,\ldots,i_k}|^2 \; ,
\end{equation}
for tree order calculations, and
\begin{equation}\sum_{\{i\},\{j\}}
\{ {\cal M}_{j_1,j_2,\ldots,j_k}^{i_1,i_2,\ldots,i_k}
{\cal L}_{j_1,j_2,\ldots,j_k}^{* i_1,i_2,\ldots,i_k}+ {\cal
L}_{j_1,j_2,\ldots,j_k}^{i_1,i_2,\ldots,i_k} {\cal
M}_{j_1,j_2,\ldots,j_k}^{* i_1,i_2,\ldots,i_k} \} \; ,
\end{equation}
for the virtual corrections, where ${\cal L}$ refers to the one-loop
amplitude.

The color sum can also be written as
\begin{equation}\label{fullcolorsum}
\sum_{\{i\},\{j\}}|{\cal
M}_{j_1,j_2,\ldots,j_k}^{i_1,i_2,\ldots,i_k}|^2=\sum_{\sigma,\sigma^\prime}A^*_{\sigma}{\cal
C}_{\sigma,\sigma^\prime}A_{\sigma^\prime} \; .
\end{equation}
The color matrix used above is given by
\begin{equation}
{\cal C}_{\sigma,\sigma\prime}\equiv \sum_{\{i\},\{j\}}
\delta_{i_{\sigma_{1}},j_1} \delta_{i_{\sigma_{2}},j_2} \ldots
\delta_{i_{\sigma_{k}},j_k} \delta_{i_{\sigma^\prime_{1}},j_1}
\delta_{i_{\sigma^\prime_{2}},j_2} \ldots
\delta_{i_{\sigma^\prime_{k}},j_k} \; ,
\end{equation}
and is equal to
\begin{equation}
{\cal C}_{\sigma,\sigma^\prime}=N_c^{m(\sigma,\sigma^\prime)} \; ,
\end{equation}
where $m(\sigma,\sigma^\prime)$ count the number of common cycles of
the two permutations, and $N_C$ is the number of colors of the
fundamental representation of the gauge group (three in our case).

Full color summation is performed in \textsc{Helac} by using the
right-hand side of Eq.~\ref{fullcolorsum}. The code generates all
possible permutations of the color indices, each of them being one color
connection. For each of the latter, using the color-connection
Feynman rules, \textsc{Helac} calculates the corresponding $A_\sigma$
color-stripped amplitude and at the end, using Eq.~\ref{fullcolorsum},
provides the fully color summed contribution.

Although at tree order, practical applications are fast enough with
full color summation, at the one-loop level, one would opt for
Monte-Carlo sampling over colors. This is done
using the left-hand side of Eq.~\ref{fullcolorsum}. The idea is rather
simple \cite{Maltoni:2002mq,Mangano:2002ea}: 
we generate a color configuration by assigning explicit
colors to the external particles. Using the labels $1,2,3$ for the
three colors in the case of QCD, a possible color assignment for a
process like $g(i_1,j_1)g(i_2,j_2)\to t(i_3,0)\bar{t}(0,j_3)
b(i_4,0)\bar{b}(0,j_4)$ can be $(i_1,\ldots,i_4)=(1,2,3,3)$ and
$(j_1,\ldots,j_4)=(3,2,1,3)$, that is ${\cal M}^{1,2,3,3}_{3,2,1,3}$
in Eq.~\ref{colorrep}. Of course, color conservation requires that
the number of times a color appears in the list of color indices
($i$), should be equal to the number of times it appears also in the
list of anti-color indices ($j$).

If one now uses Eq.~\ref{colordec}, it is easy to see that only a few
of the $A_\sigma$ amplitudes contribute to a given color assignment.
This means that including the color degree of freedom in our
Monte-Carlo integration, we may reduce drastically the average number
of color connections that are actually needed. For instance in the
calculation of $gg\to t\bar{t} b\bar{b}$, the average number of color
connections used in a Monte-Carlo sampling over color assignments, is
approximately $3$ per event, resulting to almost an order of magnitude
reduction in computation time with respect to the non sampling treatment, 
where all
color connections, $24$ in that case, have to be calculated per
event. This is particularly important for one-loop amplitudes. It
should be emphasized that Monte-Carlo over colors is not an
approximation, it is an exact treatment of the color degrees of
freedom. We have extensively tested that it produces the same results
as the usual full color sum of Eq.~\ref{fullcolorsum}, see also a previous 
work \cite{Papadopoulos:2005ky}.

The sampling over helicity is well described in 
Refs.~\cite{Draggiotis:1998gr,Draggiotis:2002hm}. It
results to a drastic improvement in speed, since only one
(random) helicity configuration has to be calculated per event,
whereas in the case for instance of $gg\to t\bar{t} b\bar{b}$ a full
summation over helicity configurations will slow down the
calculation by a factor that is approximately equal to the number of
helicity configurations, namely $32$ in that case.

Finally the actual calculation of the virtual corrections is organized
using a re-weighting technique ~\cite{Lazopoulos:2007ix,Binoth:2008kt}. 
To explain how this works, let us start with the following equation
\begin{equation}
    \sigma^{LO+V}_{ab}=\int dx_1 dx_2 d\Phi_m f_a(x_1) f_b(x_2) \left(
|{\cal M}|^2 + {\cal M}{\cal L}^*+{\cal M}^*{\cal L} \right) \; ,
\end{equation}
which gives the sum of leading order (LO) and virtual (V)
contributions for a scattering $ab\to m$-particles. It can be
re-written as
\begin{equation}
    \sigma^{LO+V}_{ab}=\int dx_1 dx_2 d\Phi_m f_a(x_1) f_b(x_2)
|{\cal M}|^2 \left(1+ \frac{{\cal M}{\cal L}^*+{\cal M}^*{\cal
L}}{|{\cal M}|^2} \right) \; .
\end{equation}
Since ${\cal L}$ is a time consuming function one would like to
calculate it as few times as possible. To this end a sample of
un-weighted events is produced based on the tree order distribution,
namely
\begin{equation}
   g(\vec{X})\equiv g(x_1,x_2,\Phi_m) = \frac{1}{\sigma^{LO}}
   \frac{d\sigma^{LO}_{ab}}{dx_1 dx_2 d\Phi_m} \; ,
\end{equation}
satisfying $\int d\vec{X} g(\vec{X})=1$. The sample $S$ of
un-weighted events has the following property,
\begin{equation}\label{rew}
   \frac{1}{N_S}\sum_{i\in S} {\cal O}(\vec{X}_i) = \int d\vec{X}
   g(\vec{X}) O(\vec{X}) \; ,
\end{equation}
where the equality should be understood in the statistical sense,
and ${\cal O}(\vec{X})$ is any well-defined function over the
integration space. Now it is trivial to see that if
\begin{equation}
   {\cal O}(\vec{X}) =\left(1+ \frac{{\cal M}{\cal L}^*+{\cal M}^*{\cal
L}}{|{\cal M}|^2} \right) \; ,
\end{equation}
then
\begin{equation}
   \frac{1}{N_S}\sum_{i\in S} {\cal O}(\vec{X}_i)
   =\frac{\sigma^{LO+V}_{ab}}{\sigma^{LO}_{ab}} \; .
\end{equation}

In practice the sample of tree order un-weighted events includes
all information on the integration space, namely, the color
assignment, the (random)helicity configuration, 
the fractions $x_1$ and $x_2$ and the $m-$body phase-space. For
future convenience it is produced in a standard Les Houches 
format \cite{Alwall:2006yp}.
One-loop contributions are only calculated for this sample of
un-weighted events, and the weight assigned to each of those events
is given by
\begin{equation}
    w=\frac{{\cal M}{\cal L}^*+{\cal M}^*{\cal L}}{|{\cal M}|^2} \; .
\end{equation}
The total virtual contribution can now be easily estimated by
\begin{equation}
    \sigma^V=\left<w\right>\sigma^B \; ,
\end{equation}
where $\sigma^B$ is the born cross section, already included in the
Les Houches file. Moreover, the sample of events including the
information on $w$, can be used to produce any kinematical
distribution, according to the Eq.~\ref{rew}.

In our application, the speed-up factor obtained in this way varies
between $10^2$ and  $10^3$. In fact, an event in the gluon-gluon channel
costs about a second on a 3 GHz machine, and few permille level cross
sections are obtained by re-weighting about fifty thousand events. The
smooth distributions of the next sections are obtained on samples of
two hundred thousand events. We stress it for the non-expert reader
that all timings are subject to drastic future improvement, but it is
easy to see that the complete calculation of the virtual corrections
is a matter of one or two days on a single machine. We consider this a
practical proof of the power of our approach.

Finally, for the expert reader this time, we stress that the numerical
stability of the virtual correction evaluation is checked by
performing a gauge independence test on each event.


\subsection{Real radiation}

The real radiation contribution to the process is obtained using the
Catani-Seyour dipole subtraction method \cite{Catani:1996vz} in the massive
version as described in \cite{Catani:2002hc} and extended for arbitrary
polarizations in \cite{Czakon:2009ss}. We use a phase space restriction on the
contribution of the dipoles as originally proposed in
\cite{Nagy:1998bb,Nagy:2003tz}. Most of the formulae needed beyond that work to
account for massive partons have been presented in
\cite{Campbell:2004ch,Campbell:2005bb}. In fact, the only missing integrated
dipoles correspond to the final emitter and final spectator case, when
both are massive. This situation occurs in our calculation, when a top
quark emits a soft gluon, which is then absorbed by the anti-top-quark
acting as the spectator of the Catani-Seymour formalism. As advertised
in the introduction, we give the complete set of expressions in the
most general case in the Appendix. Let us stress at this point, that,
similarly to most authors, we do not use finite dipoles regularizing
the quasi-collinear divergence induced by both top quarks moving in
the same direction, even though they are implemented in the software. Due to
the large top quark mass, they are not needed to improve the numerical
convergence.

At this point let us remind the reader that the phase space restriction
on the dipole phase space is defined differently depending on whether the
spectator and emitter are in the final or initial states. To be more
specific a given dipole contributes as follows
\begin{enumerate}
  
\item for final-final dipoles, if
  
  \begin{equation}
    \yijk < \alpha_{max}^{FF} \; ,
  \end{equation}
  where
  \begin{equation}
    \yijk = \frac{p_i p_j}{p_i p_j + p_i p_k + p_j p_k} \; ,
  \end{equation}
  and $p_i,p_j$ and $p_k$ are the momenta of the emitter pair and of
  the spectator respectively.
  
\item for final-initial dipoles, if
  
  \begin{equation}
    1-\xija < \alpha_{max}^{FI} \; ,
  \end{equation}
  where
  \begin{equation}
    \xija  = \frac{p_a p_i+p_a p_j-p_i p_j+\frac{1}{2}(m_{ij}^2-m_i^2-m_j^2)}
           {p_a p_i + p_a p_j} \; ,
  \end{equation}
  and $p_i,p_j$ and $p_a$ are the momenta of the emitter pair and of
  the spectator respectively. $m_{ij}$ is the emitter particle mass.
  
\item for initial-final dipoles, if
  
  \begin{equation}
    u_i < \alpha_{max}^{IF} \; ,
  \end{equation}
  where
  \begin{equation}
    u_i = \frac{p_i p_a}{p_i p_a+p_j p_a} \; ,
  \end{equation}
  and $p_a,p_i$ are the emitter pair momenta, with $p_a$ in the
  initial state, and $p_j$ is the final state spectator momentum.
  
\item for initial-initial dipoles, if
  
  \begin{equation}
    {\tilde v}_i \equiv \frac{p_a p_i}{p_a p_b} <
    \alpha_{max}^{II} \; ,
  \end{equation}
  where  $p_a,p_i$ are the emitter pair momenta, with $p_a$ in the
  initial state, and $p_b$ is the initial state spectator momentum.
  
\end{enumerate}


\begin{figure}[th]
\begin{center}
\includegraphics[width=0.44\textwidth]{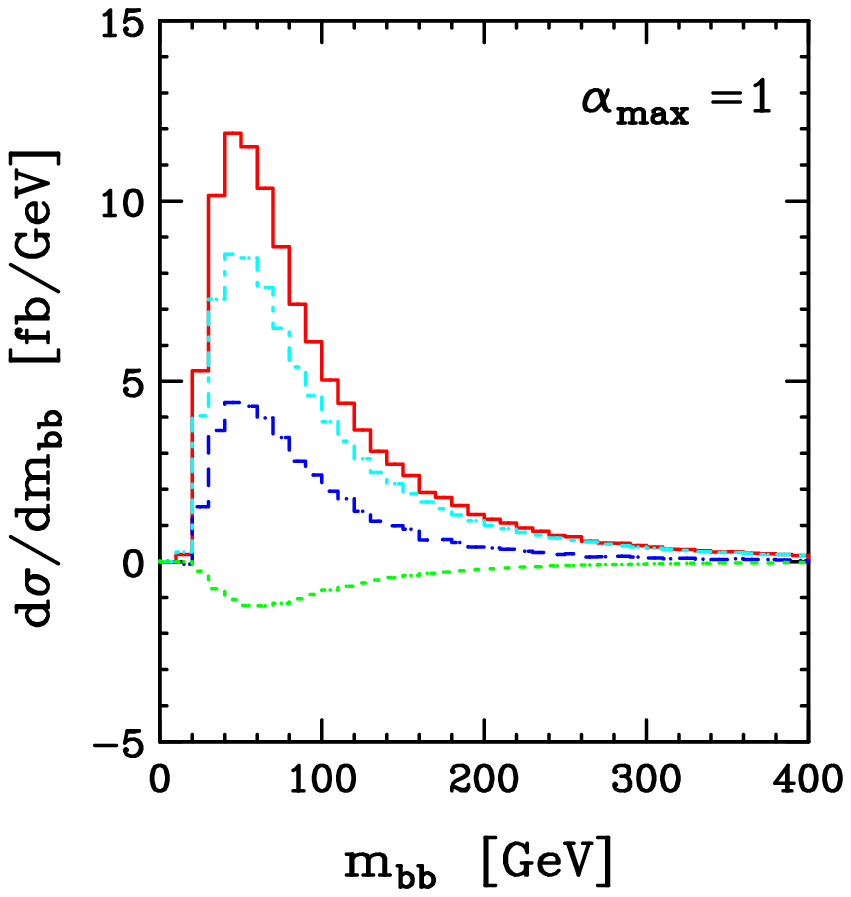}
\includegraphics[width=0.45\textwidth]{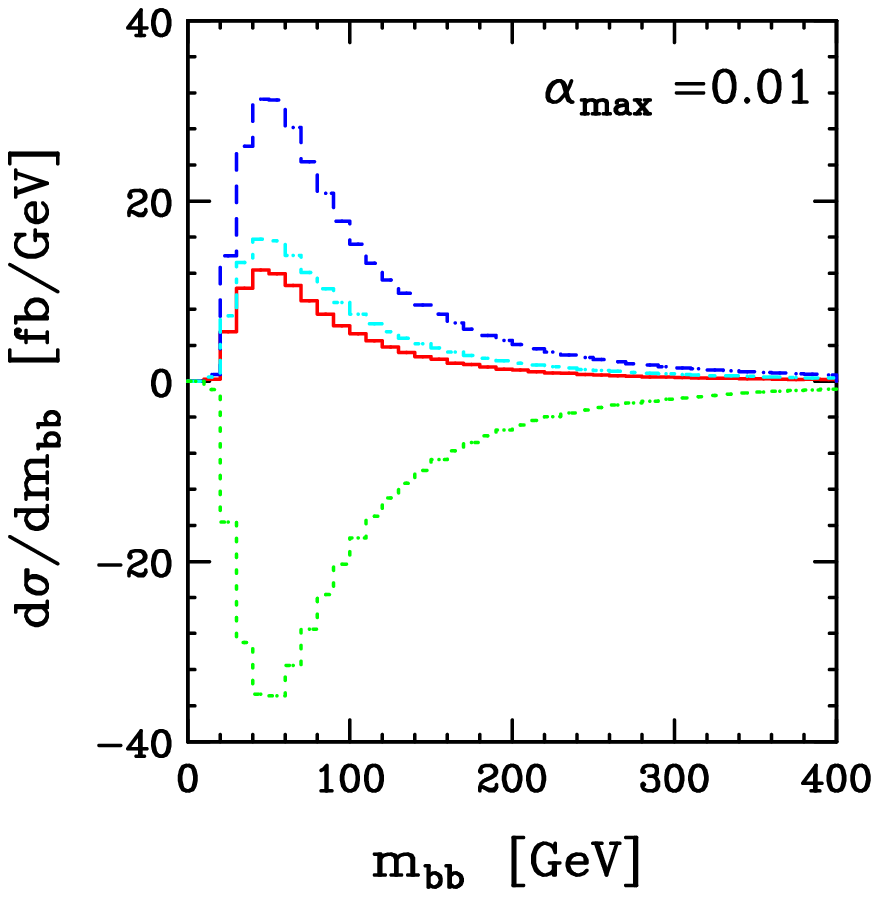}
\includegraphics[width=0.45\textwidth]{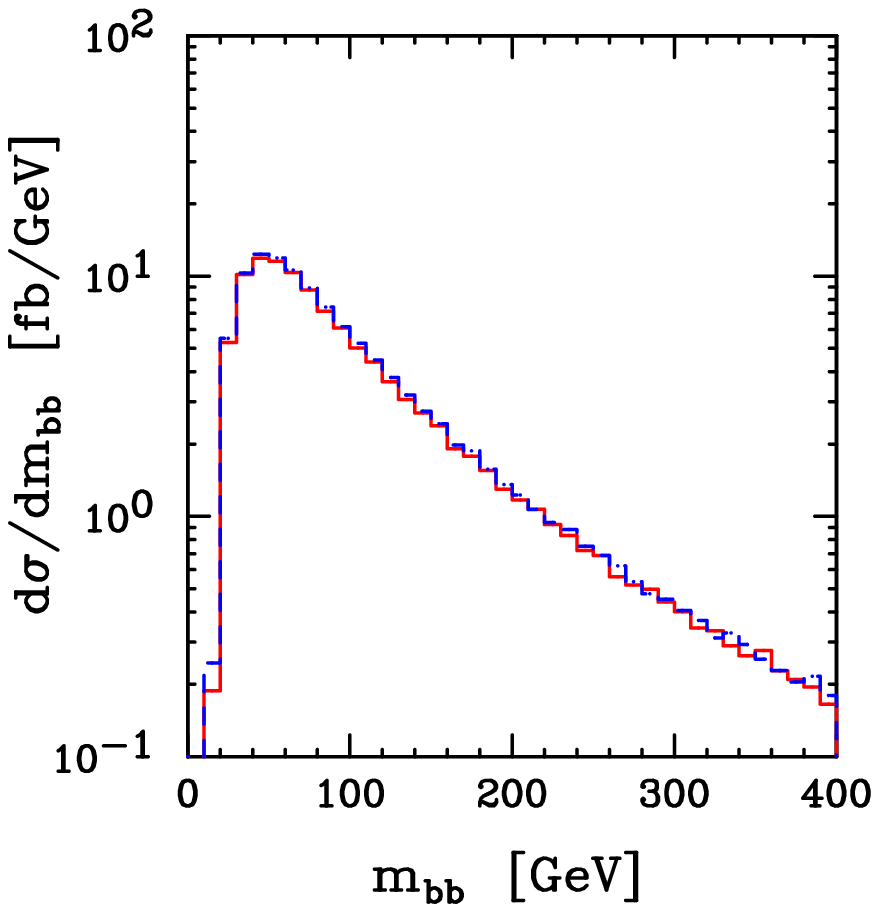}
\end{center}
\vspace{-0.2cm}
\caption{\it \label{fig:real} Distribution of the invariant mass
  $m_{b\bar{b}}$ of the bottom-anti-bottom pair for $pp\rightarrow
  t\bar{t}b\bar{b} + X$ at the LHC for different parts of the real
  radiation contribution with different choices of
  $\alpha_{max}$, $\alpha_{max} = 1$ and $\alpha_{max} = 0.01$. The red
  solid line corresponds to the sum of all contributions, the blue dashed
  line  represents the dipole subtracted real emission, the cyan
  dot-dashed line corresponds to the sum of the $K$ and $P$ insertion
  operators, and finally the green dotted line represents the $I$ insertion
  operator. The sum of all the contributions for the two different
  choices of $\alpha_{max}$ is depicted below.}
\end{figure}


In our implementation of the subtracted real radiation the $\alpha$
parameters can be varied independently of each other. On the other
hand, we slightly simplified the implementation of the integrated dipoles
by assuming $\alpha_{max}^{FF} = \alpha_{max}^{FI} =
\alpha_{max}^{IF} = \alpha_{max}^{II} \equiv \alpha_{max}$. We
consider two extreme choices, namely $\alpha_{max} = 1$ and
$\alpha_{max} = 0.01$. The first one corresponds of course to the
original formulation of \cite{Catani:2002hc}. 

Even though the use of a phase space restriction in the dipole
formalism is rather widespread currently, we wish to make a few
remarks on its advantages and disadvantages in the present
setting. On the side of advantages there are in fact three reasons
\begin{enumerate}

\item Our phase space generator, \textsc{Phegas} \cite{Papadopoulos:2000tt},
  uses multi-channel optimization \cite{Kleiss:1994qy}, 
  where the phase space density of a given
  channel corresponds to a product of Feynman diagram denominators elevated to
  some arbitrary power, which turned out to be best chosen relatively close to
  unity. This clearly reproduces the peaked behavior in the collinear and soft
  limits, with the integrable square root singularity after dipole
  subtraction. On the other hand, the distortion of the remaining
  (non-singular, but nevertheless present) peaks in the dipole subtraction
  terms due to phase space remapping is not taken into account at
  all. Therefore, restricting the kinematics to regions very close to the
  singular limits keeps the distortion small and allows to obtain a maximum
  gain from the available channels.

\item In close relation to the previous issue, we observe that there
  is more than a factor of three less events accepted if the phase space
  restriction parameter $\alpha_{max}$ is equal to $0.01$. The difference is
  due to the fact, that after phase space remapping, an event which would not
  be accepted by real radiation cuts, may pass the cuts of the dipole jet
  function if $\alpha_{max}=1$. This phenomenon is called missed-binning. It
  is important to note that despite more accepted events, the absolute error
  of the final result is only slightly better than for
  $\alpha_{max}=0.01$. In fact, to obtain the same absolute error,
  about twice less events are needed with the latter choice. This is
  is one of the two speed-up factors.

\item On the average much less dipole subtraction terms are needed per
  event with $\alpha_{max}=0.01$, since the collinear limit singles
  out a pair of partons, and the soft limit requires the sum
  over all dipoles involving the soft parton as emitter only. This constitutes
  the second speed-up factor.

\end{enumerate}

Unfortunately, having $\alpha_{max}=0.01$ implies large cancellations
between the dipole subtracted real radiation and the integrated dipole
contribution. With this choice of $\alpha_{max}$, the former
overshoots the complete result by a factor of almost three. It is
probably safe to state that a slightly larger value of $\alpha_{max}$,
such that the real radiation contribution (subtracted with dipoles)
would be similar in value to the final result for the sum of real
radiation and integrated dipoles, would be more advantageous. We leave
this issue to future studies.

It is interesting to observe the different contributions on a chosen
distribution. We illustrate such a decomposition in
Fig.~\ref{fig:real}, for the invariant mass of the $b\bar{b}$
system. All parameters and cuts are defined in the next section. 
We separate the contribution of the $I$ operator and of the
sum of the $K$ and $P$ operators defined in \cite{Catani:2002hc}. When
introducing the $\alpha_{max}$ parameter dependence, we have again
some freedom, and we choose our operators such, that the dependence of
the $I$ operator is exactly as in the massless case of \cite{Nagy:2003tz},
except for the final-final integrated dipoles for which we give the
formulae in the Appendix. Clearly, the mentioned large cancellations
for $\alpha_{max} = 0.01$ involve all three contributions, but the
integrated dipoles prove to have much better convergence and the final
error is entirely dominated by the statistically demanding real
radiation. The plot corresponding to the sum of all contributions, also
presented in Fig.~\ref{fig:real}, proves the agreement between the results for
the two parameter choices. We will show the agreement on the total cross
sections in the next section.


\section{Results}
\label{sec:results}

We consider the process $pp \rightarrow  t\bar{t} b\bar{b} + X$ at the LHC,
i.e. for  $\sqrt{s} = 14$ TeV. For the top-quark mass
we take $m_t = 172.6 $ GeV, whereas all other
QCD partons including b quarks are treated as massless. 
Mass renormalization is performed in the on-shell scheme. 
All final-state b quarks and gluons with pseudorapidity $|\eta| <5$ are 
recombined into jets with separation $\sqrt{\Delta\phi^2 +\Delta y^2} > D 
= 0.8$ in the rapidity– azimuthal-angle plane via the IR-safe $k_T$-algorithm 
\cite{Catani:1992zp,Catani:1993hr,Ellis:1993tq}. Moreover, we impose the 
following additional cuts on the transverse momenta and the rapidity of two 
recombined b-jets: $p_{T,b} > 20$ GeV, $|y_b|< 2.5$. The outgoing (anti)top
quarks are neither affected by the jet algorithm nor by phase-space cuts.  
The separation between the b-jets, $\Delta R_{b\bar{b}} > D$, implied
by the jet algorithm, together with the requirement of 
having both b-jets with $p_{T,b} > 20$ GeV sets an effective 
lower limit on the invariant mass $m_{b\bar{b}}$ 
\begin{equation}
(m_{b\bar{b}})_{min}=(p_{T,b})_{min}\sqrt{2(1-\cos D)}\approx 15.6 ~\mbox{GeV} \; ,
\end{equation}
which screens off the collinear singularity. This is the reason, why
we don't need any dipoles for the gluon splitting into a $b\bar{b}$ pair.

We consistently use the CTEQ6  set of parton distribution functions 
(PDFs) \cite{Pumplin:2002vw,Stump:2003yu} , i.e.
we take CTEQ6L1 PDFs with a 1-loop running $\alpha_s$ in LO and CTEQ6M PDFs
with a 2-loop running $\alpha_s$ in NLO, but the suppressed contribution from
b quarks in the initial state has been neglected. The number of active
flavors is $N_F = 5$, and the respective QCD parameters are $\Lambda^{LO}_5 =
165$ MeV and  $\Lambda^{MS}_5 = 226$ MeV. In the renormalization of the strong
coupling constant, the top-quark loop in the gluon self-energy is subtracted at
zero momentum. In this scheme the running of $\alpha_s$ is generated solely by
the contributions of the light-quark and gluon loops. By default, we set the 
renormalization and factorization scales, $\mu_{R}$  and $\mu_F$, to the 
common value $\mu_0 =m_t$.

We would like to stress that the above parameters correspond exactly to
those assumed in the analysis of 
\cite{Bredenstein:2008zb,Bredenstein:2009aj}, and are essentially
based on \cite{atlas}. It is clear that there are many interesting
phenomenological analyses that can be performed using our system with
different cuts, but, as explained in the introduction, our main goal
is to demonstrate its correctness and efficiency. To this end we want
to be able to compare with the previous study.

We begin our presentation of the final results of our analysis with a
discussion of the total cross section at the central value of the
scale, $\mu_R=\mu_F=m_t$. The respective numbers are presented in
Tab.~\ref{tab:comp} for the two choices of the $\alpha_{max}$
parameter. We also single out the quark channel (although its
contribution beyond leading order is negligible for any practical
study) because we can compare our results with
\cite{Bredenstein:2008zb}. Clearly, we observe perfect agreement within
statistical errors between all independent evaluations. Notice that we quote
smaller statistical errors than \cite{Bredenstein:2009aj} 
for the complete proton-proton scattering cross section, 
mostly because we performed
many experimental computations of the different sub-parts and the
accumulated statistical sample is fairly sizable. At the central scale
value, the full cross section receives a very large NLO correction of
the order of 77\% which is mainly due to the gluonic initial state as
stressed previously in \cite{Bredenstein:2009aj}.


\begin{table}[th]
\begin{center}
\begin{tabular}{c | c c | c c c}
&&&&&\\
Process & 
$\sigma^{\mbox{\footnotesize{LO}}}
_{\mbox{\footnotesize{\cite{Bredenstein:2008zb,Bredenstein:2009aj}}}}$ [fb]
& $\sigma^{\mbox{\footnotesize{LO}}}$ [fb]
& $\sigma^{\mbox{\footnotesize{NLO}}}
_{\mbox{\footnotesize{\cite{Bredenstein:2008zb,Bredenstein:2009aj}}}}$ [fb] 
& $\sigma^{\mbox{\footnotesize{NLO}}}_{\alpha_{max}=1}$ [fb]
& $\sigma^{\mbox{ \footnotesize{{NLO}}}}_{\alpha_{max}=0.01}$  [fb] \\
&&& && \\
\hline
&&&&&\\
$ q \bar{q}\rightarrow t\bar{t}b\bar{b} $  
& 85.522(26) 
& 85.489(46)
& 87.698(56)
& 87.545(91) 
& 87.581(134) \\
&&&&&\\
\hline 
&&&&&\\
$ pp\rightarrow t\bar{t}b\bar{b} $  
& 1488.8(1.2)  
& 1489.2(0.9)
& 2638(6) 
& 2642(3)
& 2636(3) \\
&&&&&\\
\end{tabular}

\caption{\it \label{tab:comp} Cross sections for $pp\rightarrow
  t\bar{t}b\bar{b} + X$ at the LHC at LO and NLO 
  for the scale choice $\mu_{F}=\mu_{R}=m_{t}$, in
  comparison with the results of 
  Refs. \cite{Bredenstein:2008zb,Bredenstein:2009aj}. 
  The statistical errors are quoted in parentheses.}
\end{center}
\end{table}


\begin{table}[th]
\begin{center}
\begin{tabular}{c | c | c | c | c | c}
&&&&&\\
 $\xi \cdot m_{t}$ 
& 1/8 $\cdot ~m_{t}$ &  1/2 $\cdot ~m_{t}$ & 
1 $\cdot ~m_{t}$ & 2 $\cdot ~m_{t}$ & 8 $\cdot ~m_{t}$ \\
&&&&&\\
\hline 
&&&&&\\
$\sigma^{\mbox{\footnotesize{LO}}}$ [fb] 
& 8885(36) 
& 2526(10)
& 1489.2(0.9)
& 923.4(3.8) 
& 388.8(1.4)\\
&&&&&\\
\hline
 &&&&&\\
$\sigma^{\mbox{\footnotesize{NLO}}}$ [fb]  
&  4213(65)
& 3498(11)
& 2636(3) 
& 1933.0(3.8) 
& 1044.7(1.7) \\
&&&&&\\
\end{tabular}
\caption{ \it \label{tab:scale} Scale dependence of the total cross
  section for $pp\rightarrow  t\bar{t}b\bar{b} + X$ at the LHC 
  at LO and NLO with $\mu_R=\mu_F=\xi m_t$ for a few distinct
  values  of $\xi$.}
 \end{center}
\end{table}


\begin{figure}[th]
\begin{center}
\includegraphics[width=0.75\textwidth]{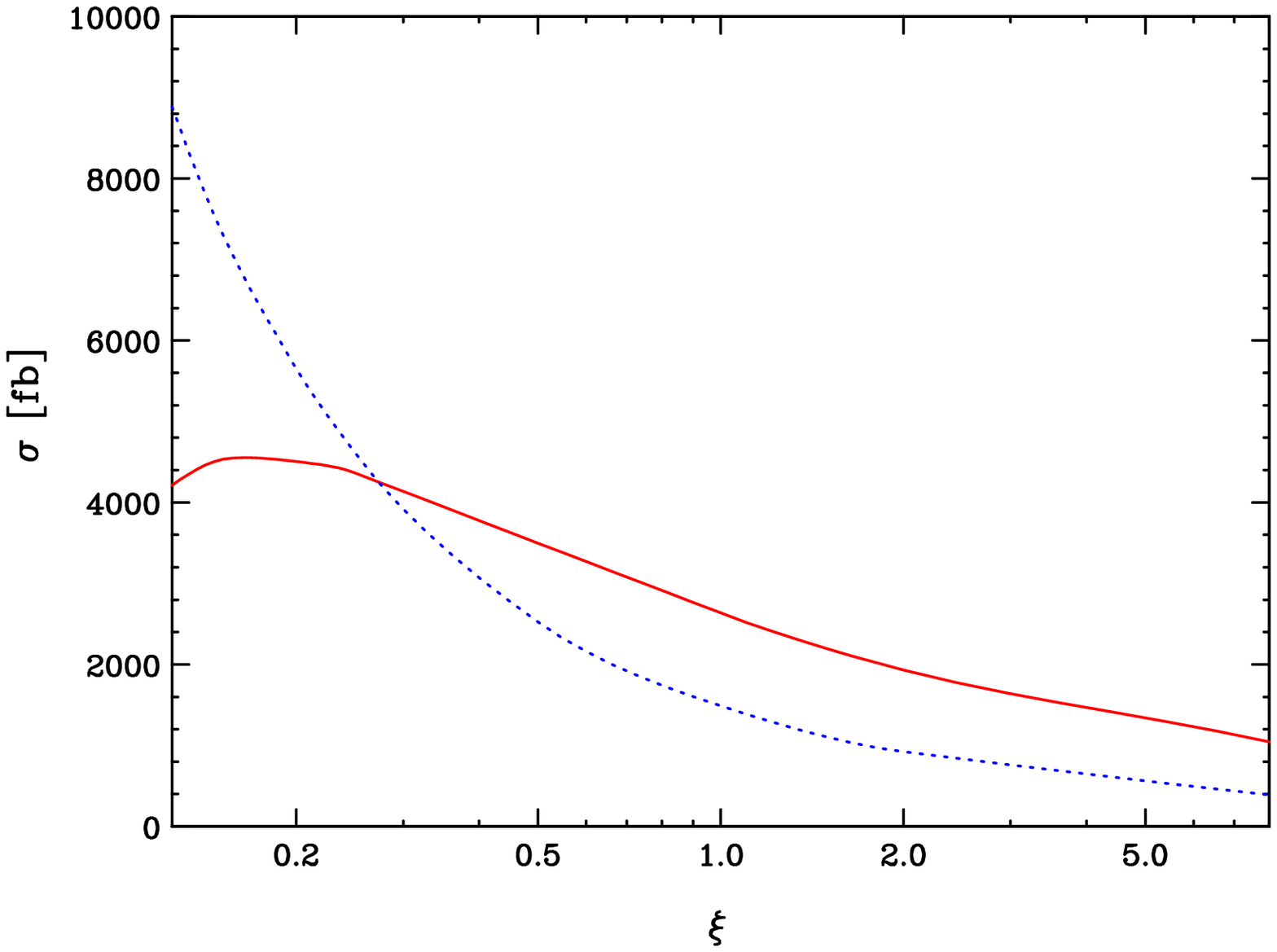}
\includegraphics[width=0.75\textwidth]{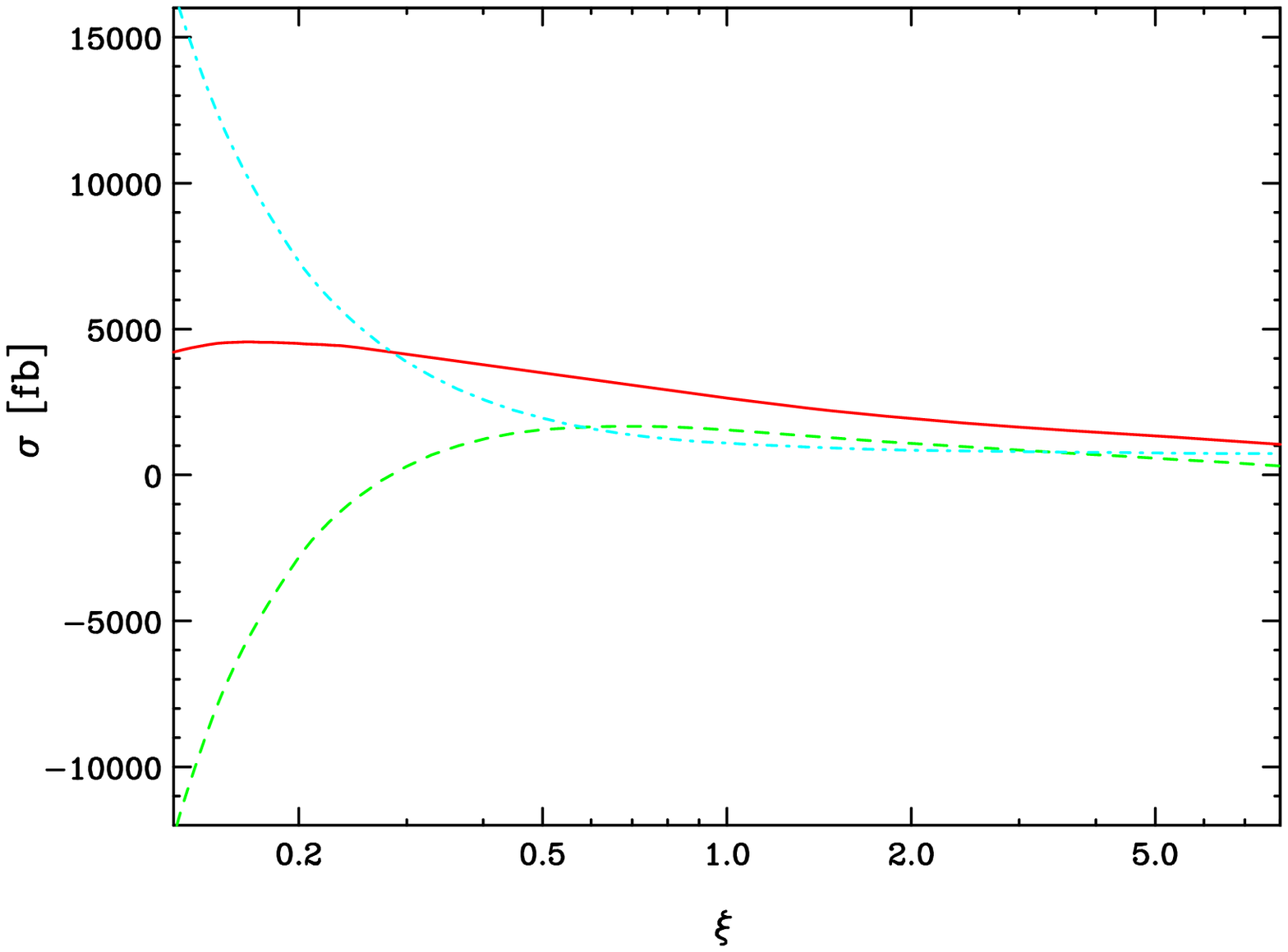}
\end{center}
\vspace{-0.2cm}
\caption{\it \label{fig:scale} Scale dependence of the total cross
  section for $pp\rightarrow  t\bar{t}b\bar{b} + X$ at the LHC  
  with $\mu_R=\mu_F=\xi m_t$. On the upper panel, the blue dashed 
  curve corresponds to the leading order, whereas the red solid one
  to the next-to-leading order result. The lower  panel shows the
  scale dependence of the next-to-leading order cross section (red solid
  curve) decomposed into the contribution of the virtual corrections
  (green dashed curved) and real radiation (cyan dash-dotted curve).}
\end{figure}


\begin{figure}[th]
\begin{center}
\includegraphics[width=0.45\textwidth]{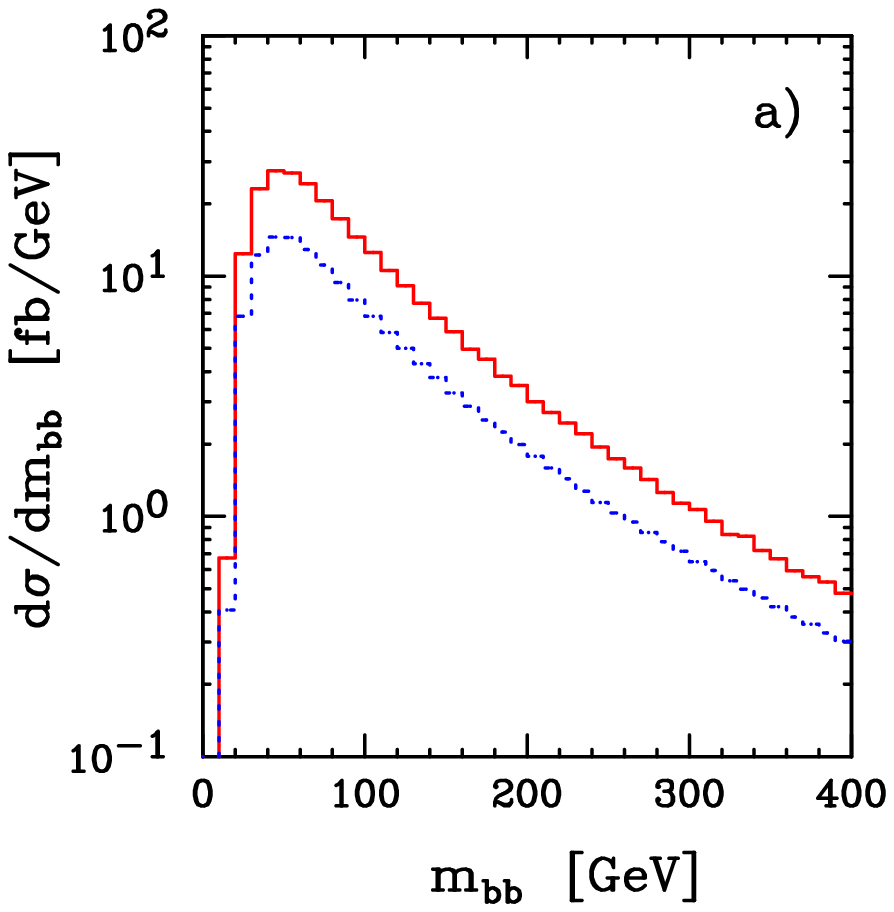}
\includegraphics[width=0.45\textwidth]{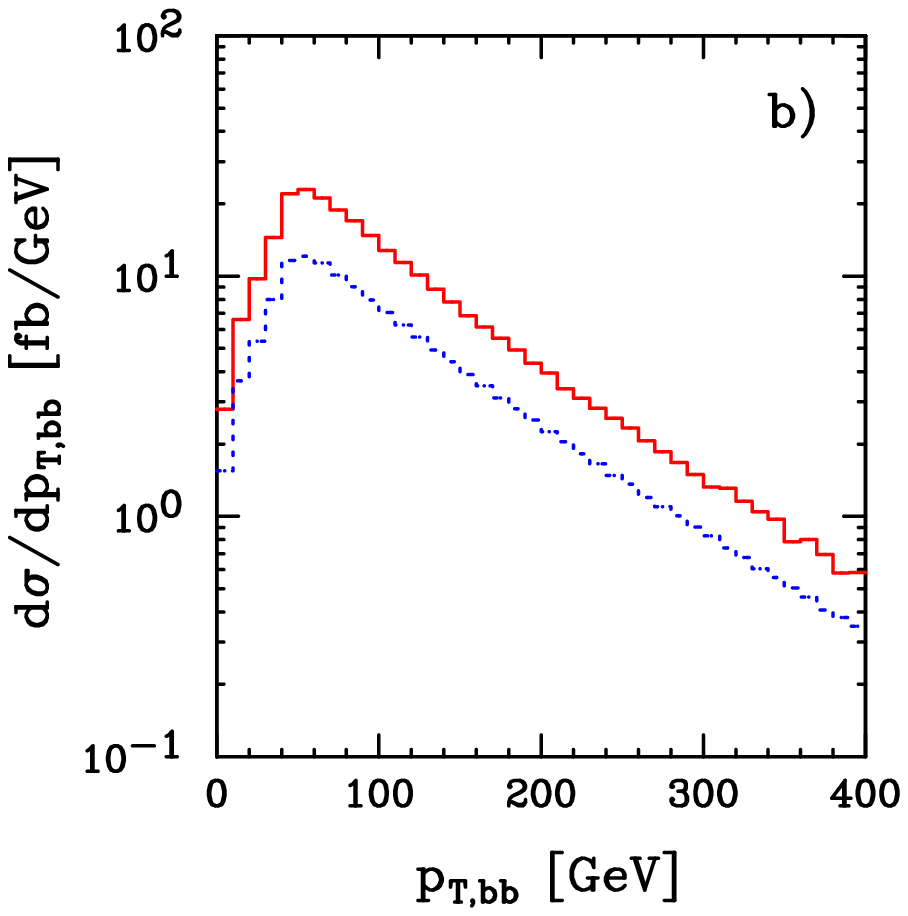}
\includegraphics[width=0.45\textwidth]{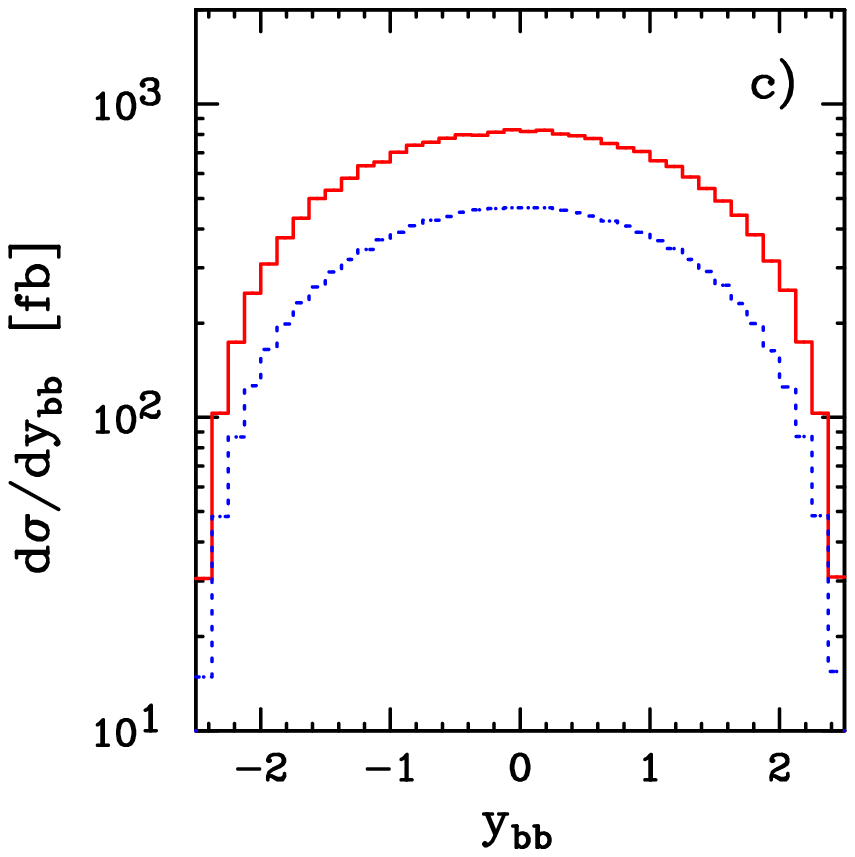}
\includegraphics[width=0.45\textwidth]{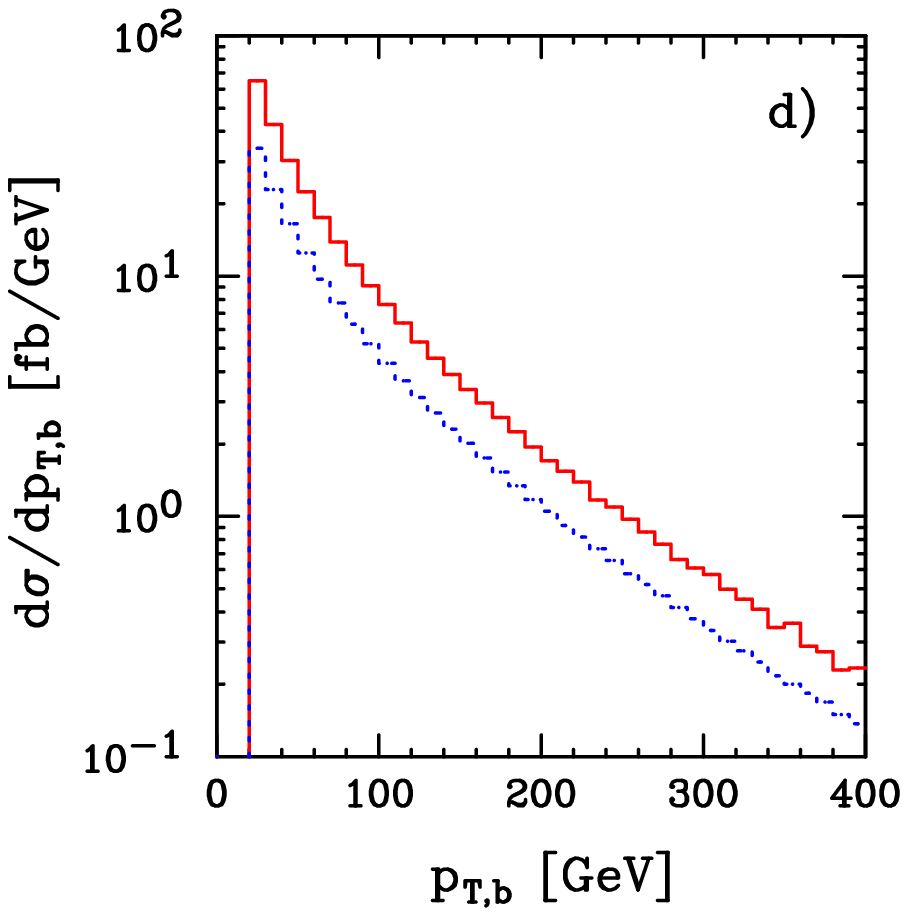}
\end{center}
\vspace{-0.2cm}
\caption{\it \label{fig:distros} Distribution of the invariant mass 
  $m_{b\bar{b}}$ of the bottom-anti-bottom pair (a), distribution in
  the transverse momentum  $p_{T_{b\bar{b}}}$  of the bottom-anti-bottom 
  pair (b), distribution in the rapidity $y_{b\bar{b}}$ of the
  bottom-anti-bottom pair (c)  and  distribution in the transverse momentum
  $p_{T_{b}}$  of the  bottom quark (d) for $pp\rightarrow t\bar{t}b\bar{b} 
  + X$ at the LHC at LO (blue dashed line) and NLO (red solid line). 
  All distributions have been obtained with $\alpha_{max}=0.01$. }
\end{figure}


\begin{figure}[t]
\begin{center}
\includegraphics[width=0.45\textwidth]{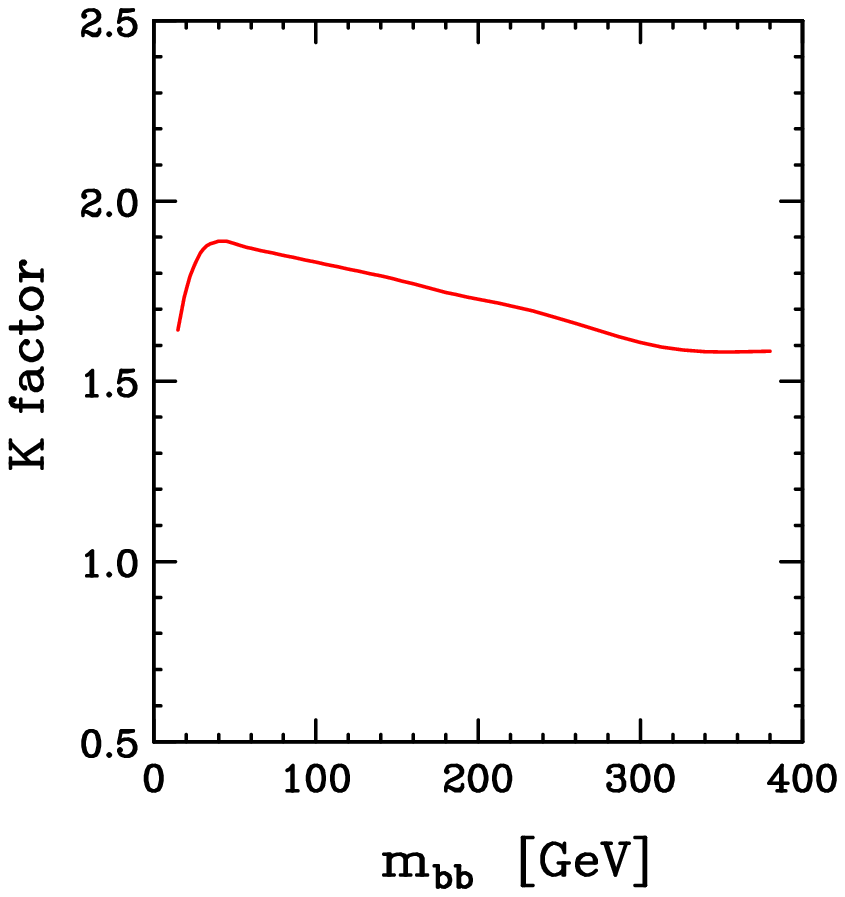}
\includegraphics[width=0.45\textwidth]{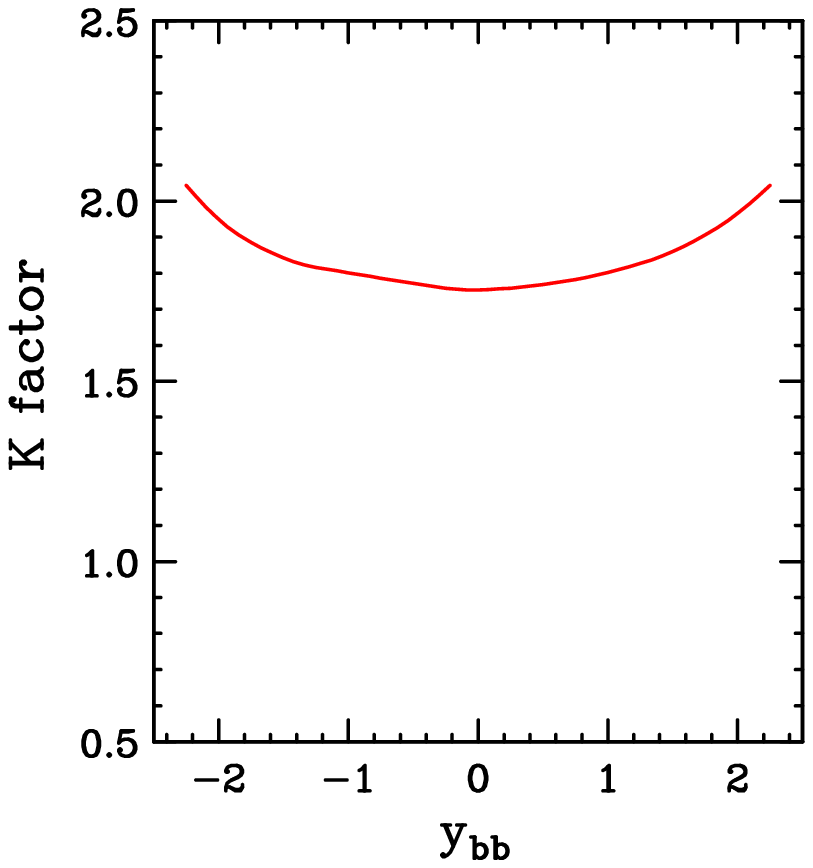}
\end{center}
\vspace{-0.2cm}
\caption{\it \label{fig:k-factors} The ratio of the NLO and LO distributions
  for $pp\rightarrow  t\bar{t}b\bar{b} + X$ at the LHC for  the invariant 
  mass  $m_{b\bar{b}}$ (left panel) and the rapidity 
  $y_{b\bar{b}}$ (right panel) of the bottom-anti-bottom pair.}
\end{figure}

Subsequently, we turn our attention to the scale dependence, which is
given in Tab.~\ref{tab:scale} for the total cross section for
$pp\rightarrow  t\bar{t}b\bar{b} + X$ at the LHC at LO and NLO with
$\mu_{\footnotesize{\mbox{R}}}=\mu_{\footnotesize{\mbox{F}}}=\xi m_t$
for a few distinct values  of $\xi$.  As expected, we observe a
reduction of the scale uncertainty while going from LO to NLO. Varying
the scale up and down by a factor 2 changes the cross section by
+70\% and -38\% in the LO case, while in the NLO case we have
obtained a variation of the order +33\% and -27\%. Our findings can be
summarized as follows
\begin{equation}
\sigma^{\mbox{\footnotesize{LO}}}_{t\bar{t}b\bar{b}}
(\mbox{LHC}, m_{t} = \mbox{176.2 GeV}, \mbox{CTEQ6L1})= 
{1489.2 ~}^{\footnotesize {\mbox{+1036.8 (70\%)}}}
_{-\footnotesize{\mbox{~~565.8 (38\%)}}} ~~\mbox{fb} \; ,
\end{equation}
\begin{equation}
\sigma^{\mbox{\footnotesize{NLO}}}_{t\bar{t}b\bar{b}}
(\mbox{LHC}, m_{t} = \mbox{176.2 GeV}, \mbox{CTEQ6M})= 
{2636 ~}^{\footnotesize {\mbox{+862 (33\%)}}}
_{-\footnotesize{\mbox{703 (27\%)}}} ~~\mbox{fb} \; .
\end{equation}

As can been easily seen from Tab.~\ref{tab:scale}, choosing a scale too
small (too large) will correspond to a very large negative (positive)
correction. Unfortunately, this large scale variation and the size of
the corrections themselves, imply that if a meaningful analysis were
required in the present setup, a full NNLO study would be
indispensable. As the latter will remain out of reach in the nearest
future, it seems that, as already suggested in \cite{Bredenstein:2009aj},
additional cuts must be introduced in order to reduce the NLO
corrections.  Only then will this process, which constitutes the main
irreducible background, not put in danger the feasibility of  Higgs boson 
searches  in the $t\bar{t}H$ channel.
     
In Fig.~\ref{fig:scale} we show the result for the scale dependence
graphically. Since we confirm the findings of 
\cite{Bredenstein:2008zb,Bredenstein:2009aj} for all
other numbers, we will not present the scale dependence with the
renormalization and factorization scales varied independently, which
can be found in that work. On the other hand it is entertaining to see
the final scale dependence for both scales equal emerge out of the two
contributions (virtual and real), as also depicted in Fig.~\ref{fig:scale}. 
Of course, the separation is entirely unphysical, but well defined once 
we state that we use what is now called 
the `t Hooft-Veltman \cite{'tHooft:1972fi} version of the dimensional 
regularization, with the integrals as defined 
in \cite{vanHameren:2009dr}. The large
cancellation for small scale values is the source of the rising
integration errors quoted in Tab.~\ref{tab:scale}.

The entirely new result, which has not been presented in the
literature until now, are the differential distributions for the four
simplest observables, namely the invariant mass, transverse momentum
and rapidity of the two-$b$-jet system, as well as the transverse
momentum of the single $b$-jet. These results can be found in
Fig.~\ref{fig:distros} and have been obtained with
$\alpha_{max}=0.01$. The histograms for $\alpha_{max}=1$ are of
similar quality, but we have refrained from averaging over the two
statistically independent evaluations. Clearly, the distributions show
the same large corrections, which turn out to be relatively constant
contrary to the case of quark initial states as shown in 
\cite{Bredenstein:2008zb}.

The histograms can also be turned into dynamical K-factors, which we
show in two cases in Fig.~\ref{fig:k-factors}, namely for the invariant
mass $m_{b\bar{b}}$  and the rapidity  $y_{b\bar{b}}$ of the
two-$b$-jet system.  For those two cases, they are simply defined as
\begin{equation}
K(m_{b\bar{b}})= \frac{d\sigma^{NLO}/dm_{b\bar{b}}}
{d\sigma^{LO}/dm_{b\bar{b}}} \; ,
\end{equation}
and
\begin{equation}
K(y_{b\bar{b}})= \frac{d\sigma^{NLO}/dy_{b\bar{b}}}
{d\sigma^{LO}/dy_{b\bar{b}}} \; ,
\end{equation}
respectively. As already anticipated above, we notice that they have a
relatively small variation, when compared with their size.


\section{Conclusions}
\label{sec:conclusions}

We have presented a complete study of the $pp\rightarrow
t\bar{t}b\bar{b} + X$ process at the LHC at the next-to
leading order in QCD. Our results agree with the previous study of
\cite{Bredenstein:2008zb,Bredenstein:2009aj}, 
for all the numbers given in that work. This fact
alone constitutes a powerful proof that our fully automatic system can
be applied in analyses of realistic processes for the Large Hadron
Collider (and for any other collider for that matter).

One conclusion, which follows from the large observed corrections, is
that in view of the importance of the process under study as background
for Higgs boson production in association with a top quark pair, a
more detailed phenomenological analysis is necessary. We postpone such
work for the future.

As a completely technical detail, our work provides the last missing
formulae for the dipole subtraction formalism with a dependence on a
dipole phase space restriction parameter. This allowed for internal
tests, as well as useful speed-ups of the calculation.


\section*{Acknowledgments}

The work of M.C. was supported by the Heisenberg Programme of
the Deutsche Forschungsgemeinschaft. G.B., C.G.P., R.P. and M.W. 
were funded in part by the RTN European Programme 
MRTN-CT-2006-035505 HEPTOOLS - Tools and
Precision Calculations for Physics Discoveries at Colliders. M.W. was
additionally supported by the
Initiative and Networking Fund of the Helmholtz Association, contract
HA-101 ("Physics at the Terascale"). 
R.P. thanks the financial support of the MEC project FPA2008-02984.


\appendix

\section{Integrated dipoles with emitter and spectator in the final state}
\label{sec:integrated}

In this Appendix, we give the formulae necessary for a complete
implementation of the phase space restriction in the dipole
subtraction formalism. We follow closely the notation of \cite{Catani:2002hc},
and present only the minimum necessary for an implementation
in a numerical program.

As stated before, the only missing ingredient concerns the case
with a final state emitter and final state spectator. The relevant
kinematical variable is
\begin{equation}
\yijk = \frac{p_i p_j}{p_i p_j + p_i p_k + p_j p_k} \; ,
\end{equation}
where $p_i,p_j$ are the momenta of the emitter pair and $p_k$ is the
momentum of the spectator. We consider the case, where $m_i^2 =
p_i^2 = 0$, and thus ignore the splitting of a gluon into a pair of
massive quarks. Note that the splitting of a heavy quark into a heavy
quark and a gluon in the presence of a massless spectator, as well as
the splitting of a massless quark in the presence of a massive
spectator have already been covered in \cite{Campbell:2005bb}. The remaining
options can be found below.

The upper limit on $\yijk$ without a phase space restriction is
\begin{equation}
y_+ = 1-\frac{2\mu_k(1-\mu_k)}{1-\mu_j^2-\mu_k^2} \; ,
\end{equation}
where $\mu_{j,k} = m_{j,k}/\sqrt{Q^2}$, with $Q = p_i+p_j+p_k$.
The restriction is imposed by adding the following condition
\begin{equation}
\yijk < \alpha \; .
\end{equation}
Clearly, we will only observe a modification if
\begin{equation}
\alpha < y_+ \; .
\end{equation}
In such a case, we write the integrated dipole function corresponding
to Eq.~5.22 of \cite{Catani:2002hc} as
\begin{equation}
I_{ij,k}(\epsilon) \rightarrow I_{ij,k}(\epsilon,\alpha) =
I_{ij,k}(\epsilon)+\Delta I_{ij,k}(\alpha) \; .
\end{equation}
Notice that, since setting a minimum value of $\yijk$ screens from all
divergences, the additional term $\Delta I_{ij,k}(\alpha)$ is finite
and can be obtained by integration over the four-dimensional phase
space, thus simplifying substantially the calculation. Indeed, we will
use the phase space of Eq.~5.11 from \cite{Catani:2002hc}, with $\epsilon=0$
and the following insertion
\begin{equation}
\int [\mbox{d} p_i(\tilde{p}_{ij},\tilde{p}_k)] \rightarrow \int
     [\mbox{d} p_i(\tilde{p}_{ij},\tilde{p}_k)] \theta(\yijk > \alpha)
     \; .
\end{equation}

We first give the result for the eikonal integral defined as
\begin{equation}
\frac{\alpha_s}{2\pi}\frac{1}{\Gamma(1-\epsilon)}
\left(\frac{4\pi\mu^2}{Q^2}\right)^\epsilon
\Delta I^{\mbox{eik}} = \int [\mbox{d} p_i(\tilde{p}_{ij},\tilde{p}_k)]
\theta(\yijk > \alpha)
\frac{1}{2p_ip_j}\frac{8\pi\mu^{2\epsilon}\alpha_s}{1-\tilde{z}_j(1-\yijk)}
\; ,
\end{equation}
with $\tilde{z}_j = p_jp_k/(p_ip_k+p_jp_k)$. It can most conveniently
be obtained by a variable change, and we therefore parametrize the
formulae with
\begin{equation}
x = y_+ - \alpha + \sqrt{(y_+ - \alpha)\left(\frac{1}{y_+}-\alpha+
\frac{4\mu_j^2\mu_k^2}{(\mu_j^2-(1-\mu_k)^2)(1-\mu_j^2-\mu_k^2)}\right)} \; .
\end{equation}
We did not attempt to minimize the number of dilogarithms and our
result contains sixteen of them
\begin{eqnarray}
\Delta I^{\mbox{eik}} =  \frac{1}{v_{jk}} && \!\!\!\!\! \left(
 - \li2\left(\frac{a + x}{a + \xp}\right)+ \li2\left(\frac{a}{a + \xp}\right)
 + \li2\left(\frac{\xp - x}{\xp - b}\right)- \li2\left(\frac{\xp}{\xp - b}\right) \right. \nonumber \\ &&
 + \li2\left(\frac{c + x}{c + \xp}\right)- \li2\left(\frac{c}{c + \xp}\right)
 + \li2\left(\frac{\xm - x}{\xm + a}\right)- \li2\left(\frac{\xm}{\xm + a}\right) \nonumber \\ &&
 - \li2\left(\frac{b - x}{b - \xm}\right)+ \li2\left(\frac{b}{b - \xm}\right)
 - \li2\left(\frac{\xm - x}{\xm + c}\right)+ \li2\left(\frac{\xm}{\xm + c}\right) \nonumber \\ &&
 + \li2\left(\frac{b - x}{b + a}\right)- \li2\left(\frac{b}{b + a}\right)
 - \li2\left(\frac{c + x}{c - a}\right)+ \li2\left(\frac{c}{c - a}\right) \nonumber \\ &&
 + \log\left(c + x\right)\log\left(\frac{(a - c)(\xp - x)}{(a + x)(c + \xp)}\right) - \log(c)\log\left(\frac{(a - c)\xp}{a(c + \xp)}\right) \nonumber \\ &&
 + \log\left(b - x\right)\log\left(\frac{(a + x)(\xm - b)}{(a + b)(\xm - x)}\right) - \log(b)\log\left(\frac{a(\xm - b)}{(a + b)\xm}\right) \nonumber \\ &&
 - \log\left((a + x)(b - \xp)\right)\log\left(\xp - x\right) + \log\left(a(b - \xp)\right)\log(\xp) \nonumber \\ &&
 + \log(d)\log\left(\frac{(a + x)\xp\xm}{a(\xp - x)(\xm-x)}\right)
 + \log\left(\frac{\xm - x}{\xm}\right)\log\left(\frac{c + \xm}{a + \xm}\right) \nonumber \\ && \left.
 + \frac{1}{2}\log\left(\frac{a + x}{a}\right)\log\left(a(a + x)(a + \xp)^2\right) \right) \; ,
\end{eqnarray}
where
\begin{eqnarray}
a &=& \frac{2\mu_k}{1 - \mu_j^2 - \mu_k^2} \; , \\
b &=& \frac{2(1 - \mu_k)}{1 - \mu_j^2 - \mu_k^2} \; , \\
c &=& \frac{2\mu_k(1 - \mu_k)}{1 - \mu_j^2 - \mu_k^2} \; ,\\
d &=& \frac{1}{2}(1 - \mu_j^2 - \mu_k^2) \; ,
\end{eqnarray}
and
\begin{eqnarray}
x_{\pm} &=& \frac{(1 - \mu_k)^2 - \mu_j^2 \pm \sqrt{\lambda(1,\mu_j^2,\mu_k^2)}}{1 - \mu_j^2 - \mu_k^2} \; , \\
v_{jk} &=& \frac{\sqrt{\lambda(1,\mu_j^2,\mu_k^2)}}{1 - \mu_j^2 - \mu_k^2} \; ,
\end{eqnarray}
with $\lambda$ the K\"allen function
\begin{equation}
\lambda(x,y,z) = x^2+y^2+z^2-2xy-2xz-2yz \; .
\end{equation}

Besides the eikonal integral we will also need the collinear
integrals (for this and future works with massive quarks), which are
implicitly defined as (in complete analogy to Eqs.~5.23-5.25 of
\cite{Catani:2002hc})
\begin{eqnarray}
\Delta I_{gQ,k} &=& C_F \left[2 \Delta I^{\mbox{eik}} + \Delta
  I^{\mbox{coll}}_{gQ,k} \right] \; , \\
\Delta I_{q\bar{q},k} &=& T_R \Delta I^{\mbox{coll}}_{q\bar{q},k} \; , \\
\Delta I_{gg,k} &=& 2 C_A \left[2 \Delta I^{\mbox{eik}} + \Delta
  I^{\mbox{coll}}_{gg,k} \right] \; .
\end{eqnarray}
The respective results read ($\mu_j$ refers to the mass of the heavy quark)
\begin{eqnarray}
\Delta I_{gQ,k} &=&  \left( \frac{3}{2}(1 + \alpha) + \frac{1}{1 - \mu_k} - 
     \frac{2(2 - 2\mu_j^2 - \mu_k)}{1 - \mu_j^2 - \mu_k^2} + 
     \frac{(1 - \alpha)\mu_j^2}
     {2(\mu_j^2 + \alpha(1 - \mu_j^2 - \mu_k^2))} \right.
     \\ && \left. - 
     2\log\left(\frac{\alpha(1 - \mu_j^2 - \mu_k^2)}{(1 - \mu_k)^2 - \mu_j^2}\right) + 
     \frac{1 + \mu_j^2 - \mu_k^2}{2(1 - \mu_j^2 - \mu_k^2)} 
        \log\left(\frac{\mu_j^2 + \alpha(1 - \mu_j^2 - \mu_k^2)}{(1 - \mu_k)^2}\right)
      \right) \; , \nonumber
\\ && \nonumber \\
\Delta I_{q\bar{q},k} &=& \frac{2}{3}\left(\frac{1 - \mu_k - \alpha(1 + \mu_k)}{1 + \mu_k} +
     \log\left(\frac{\alpha(1 + \mu_k)}{1 - \mu_k} \right) \right) \\ && +
     \left(\kappa - \frac{2}{3}\right)\frac{2\mu_k^2}{1 - \mu_k^2}\log\left(\frac{(1 -
     \alpha)(1 + \mu_k)}{2\mu_k}\right) \; , \nonumber
\\ && \nonumber \\
\Delta I_{gg,k} &=& -\frac{11}{6}\left(\frac{1 - \mu_k - \alpha(1 + \mu_k)}{1 + \mu_k} +
     \log\left(\frac{\alpha(1 + \mu_k)}{1 - \mu_k} \right) \right) \\ && -
     \left(\kappa - \frac{2}{3}\right)\frac{\mu_k^2}{1 - \mu_k^2}\log\left(\frac{(1 -
     \alpha)(1 + \mu_k)}{2\mu_k}\right) \; . \nonumber
\end{eqnarray}



\begin{thebibliography}{10}

\bibitem{Czakon:2008ii}
M.~Czakon and A.~Mitov, \href{http://arXiv.org/abs/0811.4119}
{{\tt arXiv:0811.4119 [hep-ph]}}.

\bibitem{Melnikov:2009dn}
K.~Melnikov and M.~Schulze,
{\em  JHEP} {\bf 0908} (2009) 049.

\bibitem{Czakon:2007ej}
M.~Czakon, A.~Mitov and S.~Moch, {\em Phys. Lett.} {\bf B651} (2007) 147.

\bibitem{Czakon:2007wk}
M.~Czakon, A.~Mitov and S.~Moch, {\em Nucl. Phys.} {\bf B798} (2008) 210.

\bibitem{Korner:2008bn}
J.~G. Korner, Z.~Merebashvili and M.~Rogal, {\em Phys. Rev.} {\bf D77} (2008)
  094011.

\bibitem{Czakon:2008zk}
M.~Czakon, {\em Phys. Lett.} {\bf B664} (2008) 307.

\bibitem{Bonciani:2008az}
R.~Bonciani, A.~Ferroglia, T.~Gehrmann, D.~Maitre and C.~Studerus, {\em JHEP}
  {\bf 0807} (2008) 129.

\bibitem{Anastasiou:2008vd}
C.~Anastasiou and S.~M. Aybat, {\em Phys. Rev.} {\bf D78} (2008) 114006.

\bibitem{Kniehl:2008fd}
B.~Kniehl, Z.~Merebashvili, J.~G. Korner and M.~Rogal, {\em Phys. Rev.} {\bf
  D78} (2008) 094013.

\bibitem{Bonciani:2009nb}
R.~Bonciani, A.~Ferroglia, T.~Gehrmann and C.~Studerus,
{\em  JHEP} {\bf 0908} (2009) 067.

\bibitem{Czakon:2008cx}
M.~Czakon and A.~Mitov,
 \href{http://arXiv.org/abs/0812.0353}{{\tt arXiv:0812.0353 [hep-ph]}}.

\bibitem{Beneke:2009rj}
M.~Beneke, P.~Falgari and C.~Schwinn, 
\href{http://arXiv.org/abs/0907.1443}{{\tt arXiv:0907.1443 [hep-ph]}}.

\bibitem{Czakon:2009zw}
M.~Czakon, A.~Mitov and G.~Sterman, 
\href{http://arXiv.org/abs/0907.1790}{{\tt  arXiv:0907.1790 [hep-ph]}}.

\bibitem{Beenakker:2001rj}
W.~Beenakker {\em et.~al.}, {\em Phys. Rev. Lett.} {\bf 87} (2001) 201805.

\bibitem{Reina:2001sf}
  L.~Reina and S.~Dawson, {\em  Phys. Rev. Lett.}  {\bf 87} (2001) 201804.

\bibitem{Reina:2001bc}
  L.~Reina, S.~Dawson and D.~Wackeroth, 
{\em Phys.  Rev.} {\bf D65} (2002) 053017.

\bibitem{Beenakker:2002nc}
W.~Beenakker {\em et.~al.}, {\em Nucl. Phys.} {\bf B653} (2003) 151.

\bibitem{Dawson:2002tg}
S.~Dawson, L.~H. Orr, L.~Reina and D.~Wackeroth, {\em Phys. Rev.} {\bf D67}
  (2003) 071503.

\bibitem{Dawson:2003zu}
S.~Dawson, C.~Jackson, L.~H.~Orr, L.~Reina and D.~Wackeroth, 
{\em Phys. Rev.}  {\bf D68} (2003) 034022.

\bibitem{Dittmaier:2007wz}
S.~Dittmaier, P.~Uwer and S.~Weinzierl, {\em Phys. Rev. Lett.} {\bf 98} (2007)
  262002.

\bibitem{Dittmaier:2008uj}
S.~Dittmaier, P.~Uwer and S.~Weinzierl, {\em Eur. Phys. J.} {\bf C59} (2009)
  625.

\bibitem{Lazopoulos:2008de}
A.~Lazopoulos, T.~McElmurry, K.~Melnikov and F.~Petriello, {\em Phys. Lett.}
  {\bf B666} (2008) 62.

\bibitem{Bredenstein:2008zb}
A.~Bredenstein, A.~Denner, S.~Dittmaier and S.~Pozzorini, {\em JHEP} {\bf 0808}
  (2008) 108.

\bibitem{Bredenstein:2009aj}
A.~Bredenstein, A.~Denner, S.~Dittmaier and S.~Pozzorini,
{\em  Phys.\ Rev.\ Lett.}  {\bf 103} (2009) 012002.

\bibitem{PengFei:2009ph}
D.~Peng-Fei {\em et.~al.}, \href{http://arXiv.org/abs/0907.1324}{{\tt
  arXiv:0907.1324 [hep-ph]}}.

\bibitem{atlas}
M.~Cammin, J.~Schumacher, {\tt ATL-PHYS-2003-024}.

\bibitem{Kanaki:2000ey}
A.~Kanaki and C.~G. Papadopoulos, {\em Comput. Phys. Commun.} {\bf 132} (2000)
  306.

\bibitem{Kanaki:2000ms}
A.~Kanaki and C.~G. Papadopoulos,
  \href{http://arXiv.org/abs/hep-ph/0012004}{{\tt hep-ph/0012004}}.

\bibitem{Cafarella:2007pc}
A.~Cafarella, C.~G.~Papadopoulos and M.~Worek,
 {\em  Comput. Phys. Commun.}  {\bf 180} (2009) 1941.

\bibitem{vanHameren:2009dr}
A.~van Hameren, C.~G. Papadopoulos and R.~Pittau,
  \href{http://arXiv.org/abs/0903.4665}{{\tt arXiv:0903.4665 [hep-ph]}}.

\bibitem{Ossola:2006us}
G.~Ossola, C.~G. Papadopoulos and R.~Pittau, {\em Nucl. Phys.} {\bf B763} (2007)
  147.

\bibitem{Ossola:2007ax}
G.~Ossola, C.~G. Papadopoulos and R.~Pittau, {\em JHEP} {\bf 0803} (2008) 042.

\bibitem{Czakon:2009ss}
M.~Czakon, C.~G.~Papadopoulos and M.~Worek,
{\em  JHEP} {\bf 0908} (2009) 085.

\bibitem{Draggiotis:2009yb}
P.~Draggiotis, M.~V.~Garzelli, C.~G.~Papadopoulos and R.~Pittau, 
{\em  JHEP} {\bf 0904} (2009) 072.

\bibitem{Bern:1994zx}
Z.~Bern, L.~J. Dixon, D.~C. Dunbar and D.~A. Kosower, {\em Nucl. Phys.} {\bf
  B425} (1994) 217.

\bibitem{Bern:1994cg}
Z.~Bern, L.~J. Dixon, D.~C. Dunbar and D.~A. Kosower, {\em Nucl. Phys.} {\bf
  B435} (1995) 59.

\bibitem{Witten:2003nn}
E.~Witten, {\em Commun. Math. Phys.} {\bf 252} (2004) 189.

\bibitem{Britto:2004nc}
R.~Britto, F.~Cachazo and B.~Feng, {\em Nucl. Phys.} {\bf B725} (2005) 275.

\bibitem{Brandhuber:2005jw}
A.~Brandhuber, S.~McNamara, B.~J. Spence and G.~Travaglini, {\em JHEP} 
{\bf 0510}  (2005) 011.

\bibitem{Britto:2006sj}
R.~Britto, B.~Feng and P.~Mastrolia, {\em Phys. Rev.} {\bf D73} (2006) 105004.

\bibitem{Bern:2007dw}
Z.~Bern, L.~J. Dixon and D.~A. Kosower, {\em Annals Phys.} {\bf 322} (2007)
  1587.

\bibitem{Ellis:2009zw}
  R.~K.~Ellis, K.~Melnikov and G.~Zanderighi, {\em JHEP} {\bf 0904} 
  (2009) 077.

\bibitem{Berger:2009zg}
  C.~F.~Berger {\it et al.}, {\em  Phys. Rev. Lett.}  {\bf 102} (2009) 222001.

\bibitem{Berger:2009ep}
C.~F. Berger {\em et.~al.}, \href{http://arXiv.org/abs/0907.1984}{{\tt
  arXiv:0907.1984 [hep-ph]}}.

\bibitem{Dittmaier:2007th}
S.~Dittmaier, S.~Kallweit and P.~Uwer, {\em Phys. Rev. Lett.} {\bf 100} (2008)
  062003.

\bibitem{Campbell:2007ev}
J.~M. Campbell, R.~Keith~Ellis and G.~Zanderighi, {\em JHEP} {\bf 0712} 
(2007) 056.

\bibitem{Hankele:2007sb}
V.~Hankele and D.~Zeppenfeld, {\em Phys. Lett.} {\bf B661} (2008) 103.

\bibitem{Campanario:2008yg}
F.~Campanario, V.~Hankele, C.~Oleari, S.~Prestel and D.~Zeppenfeld, {\em Phys.
  Rev.} {\bf D78} (2008) 094012.

\bibitem{FebresCordero:2008ci}
F.~Febres~Cordero, L.~Reina and D.~Wackeroth, {\em Phys. Rev.} {\bf D78} (2008)
  074014.

\bibitem{Campbell:2008hh}
J.~M.~Campbell, R.~K.~Ellis, F.~Febres Cordero, F.~Maltoni, L.~Reina, 
D.~Wackeroth and S.~Willenbrock,
{\em  Phys.\ Rev.}  {\bf D79} (2009) 034023.

\bibitem{Cordero:2009kv}
F.~F.~Cordero, L.~Reina and D.~Wackeroth,
{\em  Phys.\ Rev.}  {\bf D80} (2009) 034015.

\bibitem{Jager:2009xx}
B.~Jager, C.~Oleari and D.~Zeppenfeld,
{\em  Phys.\ Rev.} {\bf D80} (2009) 034022.

\bibitem{Bern:2008ef}
NLO Multileg Working Group Collaboration, Z.~Bern {\em et.~al.},
  \href{http://arXiv.org/abs/0803.0494}{{\tt arXiv:0803.0494 [hep-ph]}}.

\bibitem{Catani:1996vz}
S.~Catani and M.~H. Seymour, {\em Nucl. Phys.} {\bf B485} (1997) 291.

\bibitem{Catani:2002hc}
S.~Catani, S.~Dittmaier, M.~H. Seymour and Z.~Trocsanyi, {\em Nucl. Phys.} 
{\bf  B627} (2002) 189.

\bibitem{Gleisberg:2007md}
T.~Gleisberg and F.~Krauss, {\em Eur. Phys. J.} {\bf C53} (2008) 501.

\bibitem{Seymour:2008mu}
M.~H. Seymour and C.~Tevlin, \href{http://arXiv.org/abs/0803.2231}{{\tt
  arXiv:0803.2231 [hep-ph]}}.

\bibitem{Hasegawa:2008ae}
K.~Hasegawa, S.~Moch and P.~Uwer, {\em Nucl. Phys. Proc. Suppl.} {\bf 183}
  (2008) 268.

\bibitem{Frederix:2008hu}
R.~Frederix, T.~Gehrmann and N.~Greiner, {\em JHEP} {\bf 0809} (2008) 122.

\bibitem{Nagy:1998bb}
Z.~Nagy and Z.~Trocsanyi, {\em Phys. Rev.} {\bf D59} (1999) 014020.

\bibitem{Nagy:2003tz}
Z.~Nagy, {\em Phys. Rev.} {\bf D68} (2003) 094002.

\bibitem{Campbell:2004ch}
J.~M. Campbell, R.~K. Ellis and F.~Tramontano, {\em Phys. Rev.} {\bf D70} (2004)
  094012.

\bibitem{Campbell:2005bb}
J.~M. Campbell and F.~Tramontano, {\em Nucl. Phys.} {\bf B726} (2005) 109.

\bibitem{Maltoni:2002mq}
F.~Maltoni, K.~Paul, T.~Stelzer and S.~Willenbrock, {\em Phys. Rev.} {\bf D67}
  (2003) 014026.

\bibitem{Mangano:2002ea}
M.~L. Mangano, M.~Moretti, F.~Piccinini, R.~Pittau and A.~D. Polosa, {\em JHEP}
  {\bf 0307} (2003) 001.

\bibitem{Papadopoulos:2005ky}
C.~G. Papadopoulos and M.~Worek, {\em Eur. Phys. J.} {\bf C50} (2007) 843.

\bibitem{Draggiotis:1998gr}
P.~Draggiotis, R.~H.~P. Kleiss and C.~G. Papadopoulos, {\em Phys. Lett.} {\bf
  B439} (1998) 157.

\bibitem{Draggiotis:2002hm}
P.~D. Draggiotis, R.~H.~P. Kleiss and C.~G. Papadopoulos, {\em Eur. Phys. J.}
  {\bf C24} (2002) 447.

\bibitem{Lazopoulos:2007ix}
A.~Lazopoulos, K.~Melnikov and F.~Petriello, {\em Phys. Rev.} {\bf D76} (2007)
  014001.

\bibitem{Binoth:2008kt}
T.~Binoth, G.~Ossola, C.~G. Papadopoulos and R.~Pittau, {\em JHEP} {\bf 0806}
  (2008) 082.

\bibitem{Alwall:2006yp}
J.~Alwall {\em et.~al.}, {\em Comput. Phys. Commun.} {\bf 176} (2007) 300.

\bibitem{Papadopoulos:2000tt}
C.~G. Papadopoulos, {\em Comput. Phys. Commun.} {\bf 137} (2001) 247.

\bibitem{Kleiss:1994qy}
R.~Kleiss and R.~Pittau, {\em Comput. Phys. Commun.} {\bf 83} (1994) 141.

\bibitem{Catani:1992zp}
S.~Catani, Y.~L. Dokshitzer and B.~R. Webber, {\em Phys. Lett.} {\bf B285}
  (1992) 291.

\bibitem{Catani:1993hr}
S.~Catani, Y.~L. Dokshitzer, M.~H. Seymour and B.~R. Webber, {\em Nucl. Phys.}
  {\bf B406} (1993) 187.

\bibitem{Ellis:1993tq}
S.~D. Ellis and D.~E. Soper, {\em Phys. Rev.} {\bf D48} (1993) 3160.

\bibitem{Pumplin:2002vw}
J.~Pumplin {\em et.~al.}, {\em JHEP} {\bf 0207} (2002) 012.

\bibitem{Stump:2003yu}
D.~Stump {\em et.~al.}, {\em JHEP} {\bf 0310} (2003) 046.

\bibitem{'tHooft:1972fi}
G.~'t~Hooft and M.~J.~G. Veltman, {\em Nucl. Phys.} {\bf B44} (1972) 189.

\end{thebibliography}

\providecommand{\href}[2]{#2}\begingroup\raggedright\endgroup

\end{document}